# Near-field radiative heat transfer between arbitrarily-shaped objects and a surface

Sheila Edalatpour and Mathieu Francoeur[†]

*Radiative Energy Transfer Lab, Department of Mechanical Engineering*

*University of Utah, Salt Lake City, UT 84112, USA*

## ABSTRACT

A fluctuational electrodynamics-based formalism for calculating near-field radiative heat transfer between objects of arbitrary size and shape and an infinite surface is presented. The surface interactions are treated analytically via Sommerfeld's theory of electric dipole radiation above an infinite plane. The volume integral equation for the electric field is discretized using the thermal discrete dipole approximation (T-DDA). The framework is verified against exact results in the sphere-surface configuration, and is applied to analyze near-field radiative heat transfer between a complex-shaped probe and an infinite plane both made of silica. It is found that when the probe tip size is approximately equal to or smaller than the gap $d$ separating the probe and the surface, coupled localized surface phonon (LSPh)-surface phonon-polariton (SPhP) mediated heat transfer occurs. In this regime, the net spectral heat rate exhibits four resonant modes due to LSPhs along the minor axis of the probe while the net total heat rate in the near field follows a $d^{-0.3}$ power law. Conversely, when the probe tip size is much larger than the separation gap $d$, heat transfer is mediated by SPhPs resulting in two resonant modes in the net spectral heat rate corresponding to those of a single emitting silica surface while the net total heat rate approaches a $d^{-2}$ power law. It is also demonstrated that a complex-shaped probe can be approximated by a

---

[†] Corresponding author. Tel.: +1 801 581 5721, Fax: +1 801 585 9825

E-mail address: mfrancoeur@mech.utah.edu

prolate spheroidal electric dipole when the thermal wavelength is larger than the major axis of the spheroidal dipole and when the separation gap *d* is much larger than the radius of curvature of the dipole tip facing the surface.

## I. INTRODUCTION

Near-field radiative heat transfer between arbitrarily-shaped objects and a surface is of importance in many engineering applications such as near-field thermal spectroscopy and imaging [1-4], tip-based nanomanufacturing [5-7] and localized radiative cooling [8]. An analytical solution for this type of problem is only available for the case of a single sphere above an infinite plane [9,10]. An expression for the radiative heat flux between two arbitrarily-shaped objects, including surfaces, written in terms of reflection and transmission matrices has been derived using a scattering-matrix approach [11,12]. However, an explicit closed-form solution for the radiative heat flux has been provided only for the case of two slabs [12]. Simplified formulations, namely the proximity and electric dipole approximations, have been used to model experiments involving a micro/nanosized object and a large surface exchanging thermal radiation [1,3,4,13-16]. The proximity approximation is valid when the object size is much larger than its distance relative to the surface and when the object is optically thick [17,18]. When these conditions are satisfied, the heat rate between the object and the surface can be modeled as a summation of local heat rates between two parallel planes [15]. The electric dipole approximation is valid when the size of the object is much smaller than the thermal wavelength and its distance relative to the surface. Various electric dipole formulations have been proposed for modeling near-field thermal interactions between a small object and a surface. These formulations include a spherical dipole above a flat [19] and a structured [20] surface, a spherical dipole with dressed polarizability above a flat surface [16], and a spheroidal dipole above a flat surface [21]. An upper limit for the near-field radiative heat transfer between a dipole and a surface has been derived in Ref. [22]. These simplified models are however valid under limiting conditions that are often not satisfied in actual experiments. Accurate modeling of near-field radiative heat transfer between arbitrarily-shaped objects and a surface that does not rely on simplified formulations and fitting parameters can be done via numerical methods. So far, the finite-difference frequency-domain (FDFD) method [23], the finite-difference time-domain (FDTD) method [24,25], the boundary element method (BEM) [26], the thermal discrete dipole approximation (T-DDA) [18,27] and a fluctuating volume-current method [28] have been applied to numerical simulation of near-field thermal radiation problems. The FDFD and FDTD are based on discretizing the differential form of Maxwell's equations. These methods are typically computationally expensive as they require discretization of the free space in addition to the objects. In the BEM, the surface integral form of Maxwell's equations is discretized such that this approach is difficult to apply to inhomogeneous media. The T-DDA and the fluctuating volume-current method are based on discretizing the volume integral form of Maxwell's equations and can thus handle easily inhomogeneous media. Yet, these numerical approaches are difficult to apply to multi-scale problems involving a surface and micro/nanosized objects due to the prohibitive calculation time associated with discretizing a surface that is many orders of magnitude larger than the objects. The only numerical formulation capable of handling nontrivial geometries and an infinite surface is a combination of a scattering-based approach and the BEM [29].

In this paper, a framework for modeling near-field radiative heat transfer between objects and an infinite surface is provided. The formalism, based on fluctuational electrodynamics [30], is independent of the size, shape and number of objects. The volume integral equation for the electric field derived from fluctuational electrodynamics is discretized using the thermal discrete dipole approximation (T-DDA) [18,27,31]. The interactions between the objects and the surface are treated analytically using Sommerfeld's theory of electric dipole radiation above an infinite plane [32]. This approach, also used in the discrete dipole approximation literature for predicting light scattering by particles on or near a surface [33-40], does not necessitate discretization of the surface. The T-DDA with surface interaction is afterwards applied to study near-field radiative heat transfer between a probe and a surface. Understanding the thermal interactions in the probe-surface configuration is of interest in near-field thermal spectroscopy where two independent experimental studies reported resonance redshift of the scattered thermal near field [3,4]. McCauley et al. [29] analyzed the total heat rate between a conically-shaped probe and an infinite surface as well as the spatial distribution of power absorbed within the surface. The total heat rate between a cone and a disk was also modeled by Rodriguez et al. [41]. Kim et al. [42] investigated the validity of fluctuational electrodynamics in the extreme near field by measuring the heat rate between a dull probe and a surface. The probing tip was modeled as a hemisphere for which the heat rate could also be obtained via the proximity approximation. Kloppstech et al. [43] modeled the total heat rate between a conically-shaped probe with a hemispherical tip and a finite-sized surface for comparison against near-field scanning thermal microscope measurements. In this work, the spectral and total heat rate between a probe with a tip size smaller than, approximately equal to, and larger than the separation gap is studied for the first time. The validity of the spheroidal electric dipole approximation for predicting near-field radiative heat transfer between a probe and a surface is also discussed.

The paper is organized as follows. The framework for calculating near-field radiative heat transfer between arbitrarily-shaped objects and a surface is presented in Section II. In Section III, the T-DDA with surface interaction is verified against the exact solution of heat rate between a sphere and a surface. Near-field radiative heat transfer between a probe and a surface is analyzed in Section IV. Concluding remarks are then provided.

## II. DESCRIPTION OF THE FRAMEWORK

### A. Volume integral equation for the electric field

The formalism described hereafter is based on fluctuational electrodynamics, and is thus valid for heat sources in local thermodynamic equilibrium. The problem under consideration is shown in Fig. 1, where radiative heat transfer between objects submerged in vacuum and an infinite surface is to be calculated. The vacuum, the surface and the objects are referred to as medium 0, 1 and 2, respectively. It is assumed that the objects of arbitrary number, shape and size occupy a total volume $V_2$ and are isotropic, linear and nonmagnetic. Individual objects may have different, inhomogeneous temperatures $T_2$ and frequency-dependent dielectric functions $\varepsilon_2$ local in space. The surface, of volume $V_1$ and uniform temperature $T_1$, is assumed to be isotropic, linear, nonmagnetic and is characterized by a homogeneous dielectric function $\varepsilon_1$ local in space. Illumination by external sources such as laser irradiation or thermal emission by the surroundings (i.e., the thermal bath) is modeled via an incident electric field $\mathbf{E}^{inc}$. The incident electric field can originate from above or below the surface. The electric field thermally emitted by the surface into the vacuum of volume $V_0$ is denoted by $\mathbf{E}^{sur}$.

The net radiative heat rate between the objects and the surface is derived from the stochastic Maxwell equations, where a fluctuating current $\mathbf{J}^{fl}$ representing thermal emission is added to Ampère's law [30]. The ensemble average of the fluctuating current is zero, while the ensemble average of the spatial correlation function of the fluctuating current is related to the local temperature of a heat source via the fluctuation-dissipation theorem [30]:

$$\langle \mathbf{J}^{fl}(\mathbf{r}',\omega) \otimes \mathbf{J}^{fl}(\mathbf{r}'',\omega') \rangle = \frac{4\omega\varepsilon_0\varepsilon''}{\pi}\Theta(\omega,T)\delta(\mathbf{r}'-\mathbf{r}'')\delta(\omega-\omega')\bar{\bar{\mathbf{I}}} \qquad (1)$$

where $\otimes$ denotes the outer product, $\bar{\bar{\mathbf{I}}}$ is the unit dyadic, $\varepsilon_0$ is the electric permittivity of vacuum, $\varepsilon''$ is the imaginary part of the dielectric function of the heat source and $\Theta(\omega,T)$ is the mean energy of an electromagnetic state given by $\Theta(\omega,T) = \hbar\omega / [\exp(\hbar\omega/k_B T) - 1]$.

The electric field everywhere above the surface satisfies the following vector wave equation derived from the stochastic Maxwell equations [18,31]:

$$\nabla \times \nabla \times \mathbf{E}(\mathbf{r},\omega) - k_0^2 \mathbf{E}(\mathbf{r},\omega) = i\omega\mu_0 \mathbf{J}(\mathbf{r},\omega), \quad \mathbf{r} \in V_0 \cup V_2 \qquad (2)$$

where $k_0$ and $\mu_0$ are the magnitude of the wavevector and the magnetic permeability of vacuum, respectively, $i$ is the complex constant and $\mathbf{r}$ is the position vector where the fields are observed in $V_0 \cup V_2$. The current **J** in Eq. (2) is an equivalent source function generating fluctuating and scattered electric fields:

$$\mathbf{J}(\mathbf{r},\omega) = \mathbf{J}_2^{fl}(\mathbf{r},\omega) - i\omega\varepsilon_0(\varepsilon_2(\mathbf{r}) - 1)\mathbf{E}(\mathbf{r},\omega), \quad \mathbf{r} \in V_2 \qquad (3)$$

where the subscript 2 in $\mathbf{J}_2^{fl}$ specifies that the fluctuating current is in $V_2$. Note that the current **J** vanishes in $V_0$.

The solution of the inhomogeneous linear differential equation (2) is the sum of the solution of the homogeneous equation and a particular solution of the inhomogeneous equation. The homogeneous vector wave equation is given by:

$$\nabla \times \nabla \times \left(\mathbf{E}^{inc}(\mathbf{r},\omega) + \mathbf{E}^{sur}(\mathbf{r},\omega)\right) - k_0^2 \left(\mathbf{E}^{inc}(\mathbf{r},\omega) + \mathbf{E}^{sur}(\mathbf{r},\omega)\right) = \mathbf{0}, \quad \mathbf{r} \in V_0 \cup V_2 \tag{4}$$

The solution of Eq. (4) provides the electric field that would exist above the surface in the absence of objects. This electric field is comprised of two components, namely the incident field $\mathbf{E}^{inc}$ and the surface field $\mathbf{E}^{sur}$. The surface field is generated by fluctuating currents in $V_1$, $\mathbf{J}_1^{fl}$, and its expression is given by:

$$\mathbf{E}^{sur}(\mathbf{r},\omega) = i\omega\mu_0 \int_{V_1} \overline{\overline{\mathbf{G}}}^T(\mathbf{r},\mathbf{r}',\omega) \cdot \mathbf{J}_1^{fl}(\mathbf{r}',\omega) d^3\mathbf{r}', \quad \mathbf{r} \in V_0 \cup V_2 \tag{5}$$

where $\overline{\overline{\mathbf{G}}}^T$ is the transmission dyadic Green's function (DGF) relating the field observed at $\mathbf{r}$ in $V_0 \cup V_2$ to a source point $\mathbf{r}'$ located in $V_1$ [44,45]. The expression for the incident field must satisfy Eq. (4) and depends on the external radiation source.

The particular solution of Eq. (2) is the sum of the fluctuating and scattered electric fields generated by the current $\mathbf{J}$. The fluctuating and scattered fields are obtained using DGFs relating the electric field observed at $\mathbf{r}$ to a source located at $\mathbf{r}'$, as shown in Fig. 2, when both $\mathbf{r}$ and $\mathbf{r}'$ are located above the surface in $V_0 \cup V_2$:

$$\mathbf{E}^{sca}(\mathbf{r},\omega) + \mathbf{E}^{fl}(\mathbf{r},\omega) = i\omega\mu_0 \int_{V_2} \overline{\overline{\mathbf{G}}}(\mathbf{r},\mathbf{r}',\omega) \cdot \mathbf{J}(\mathbf{r}',\omega) d^3\mathbf{r}', \quad \mathbf{r} \in V_0 \cup V_2 \tag{6}$$

The DGF $\overline{\overline{\mathbf{G}}}$ is comprised of two components. The first component is the free space DGF, $\overline{\overline{\mathbf{G}}}^0$, that accounts for the electric field generated at **r** due to direct radiation by the source **J** located at **r**′ in the absence of the surface. The second component is the reflection DGF, $\overline{\overline{\mathbf{G}}}^R$, representing the electric field generated at **r** due to radiation by the source **J** located at **r**′ after reflection by the surface.

The volume integral equation for the total electric field in $V_0 \cup V_2$ is obtained by adding the incident and surface fields to Eq. (6):

$$\mathbf{E}(\mathbf{r},\omega) = i\omega\mu_0 \int_{V_2} \left( \overline{\overline{\mathbf{G}}}^0(\mathbf{r},\mathbf{r}',\omega) + \overline{\overline{\mathbf{G}}}^R(\mathbf{r},\mathbf{r}',\omega) \right) \cdot \mathbf{J}(\mathbf{r}',\omega) d^3\mathbf{r}' \\ + \mathbf{E}^{inc}(\mathbf{r},\omega) + \mathbf{E}^{sur}(\mathbf{r},\omega), \quad \mathbf{r} \in V_0 \cup V_2 \tag{7}$$

The magnetic field in $V_0 \cup V_2$ can be obtained from Eq. (7) using Faraday's law.

The solution of Eq. (7) provides the electric field in $V_2$ from which heat transfer is calculated. An analytical solution of Eq. (7) only exists for a single sphere above an infinite surface [9,10]. For arbitrarily-shaped objects, numerical approaches should be considered. Here, the T-DDA [18,27,31] is used for solving Eq. (7) and thus for computing radiation heat transfer.

**B. Radiative heat transfer calculations with the thermal discrete dipole approximation (T-DDA)**

The T-DDA formulation is initiated by discretizing $V_2$ into $N$ cubical subvolumes. The size of the subvolumes must be smaller than all characteristic lengths of the problem, namely the wavelength in $V_2$ and vacuum as well as the object-object and object-surface separation distances. In addition, the subvolume size must be small enough to represent accurately the object shape via a cubical lattice. When these conditions are fulfilled, the electric field, the DGFs and the electromagnetic properties can be assumed uniform inside a given subvolume. Under the approximation of uniform electric field, it is possible to conceptualize the subvolumes as electric point dipoles. The total dipole moment associated with a subvolume $i$ of volume $\Delta V_i$ is related to the equivalent current via $\mathbf{p}_i = (i/\omega)\int_{\Delta V_i} \mathbf{J}(\mathbf{r}',\omega)d^3\mathbf{r}'$. The discretized volume integral equation for the electric field (7) can thus written in terms of dipole moments as follows:

$$\frac{1}{\alpha_i}\mathbf{p}_i - \frac{k_0^2}{\varepsilon_0}\sum_{j\neq i}\overline{\overline{\mathbf{G}}}_{ij}^{0}\cdot\mathbf{p}_j - \frac{k_0^2}{\varepsilon_0}\sum_{j}\overline{\overline{\mathbf{G}}}_{ij}^{R}\cdot\mathbf{p}_j = \frac{3}{(\varepsilon_{2,i}+2)}\frac{1}{\alpha_i^{CM}}\mathbf{p}_i^{fl} + \mathbf{E}_i^{inc} + \mathbf{E}_i^{sur} \qquad (8)$$

where the DGFs $\overline{\overline{\mathbf{G}}}_{ij}^{0}$ and $\overline{\overline{\mathbf{G}}}_{ij}^{R}$ are evaluated between the center points of subvolumes $i$ and $j$. The total dipole moment $\mathbf{p}_i$ is the sum of two contributions, namely an induced dipole moment $\mathbf{p}_i^{ind} = \Delta V_i \varepsilon_0 (\varepsilon_{2,i}-1)\mathbf{E}_i$ and a thermally fluctuating dipole moment $\mathbf{p}_i^{fl} = (i/\omega)\int_{\Delta V_i}\mathbf{J}_2^{fl}(\mathbf{r}',\omega)d^3\mathbf{r}'$. Using this last expression in combination with the fluctuation-dissipation theorem (1), the ensemble average of the spatial correlation function of fluctuating dipole moments can be expressed in terms of the local temperature of the medium [18]. The terms $\alpha_i^{CM}$ and $\alpha_i$ are the Clausius-Mossotti and radiative polarizabilities given by:

$$\alpha_i^{CM} = 3\varepsilon_0 \Delta V_i \frac{\varepsilon_{2,i}-1}{\varepsilon_{2,i}+2} \qquad (9)$$

$$\alpha_i = \frac{\alpha_i^{CM}}{1-(\alpha_i^{CM}/2\pi\varepsilon_0 a_i^3)[e^{ik_0 a_i}(1-ik_0 a_i)-1]} \qquad (10)$$

where $a_i$ is the radius of a sphere of volume $\Delta V_i$. The application of Eq. (8) to all $N$ subvolumes in $V_2$ results in a system of equations that can be written in a matrix form:

$$\left(\overline{\overline{\mathbf{A}}}+\overline{\overline{\mathbf{R}}}\right)\cdot\overline{\mathbf{P}} = \overline{\mathbf{E}}^{fdt} + \overline{\mathbf{E}}^{inc} + \overline{\mathbf{E}}^{sur} \tag{11}$$

where $\overline{\mathbf{E}}^{fdt}$ and $\overline{\mathbf{E}}^{sur}$ are $3N$ stochastic column vectors containing the first term on the right-hand side of Eq. (8) and the surface field, respectively, while $\overline{\mathbf{E}}^{inc}$ is the $3N$ deterministic column vector containing the incident field. The term $\overline{\mathbf{P}}$ is the $3N$ stochastic column vector containing the unknown total dipole moments of the subvolumes. The matrix $\overline{\overline{\mathbf{A}}}$ is the $3N$ by $3N$ deterministic interaction matrix which is composed of submatrices $\overline{\overline{\mathbf{A}}}_{ij}$ representing the direct interaction between subvolumes $i$ and $j$ in the absence of the surface. A submatrix $\overline{\overline{\mathbf{A}}}_{ij}$, obtained from the free space DGF, is calculated using the following expressions when $i \neq j$:

$$\overline{\overline{\mathbf{A}}}_{ij} = C_{ij} \begin{bmatrix} \beta_{ij}+\gamma_{ij}\hat{r}_{ij,x}^2 & \gamma_{ij}\hat{r}_{ij,x}\hat{r}_{ij,y} & \gamma_{ij}\hat{r}_{ij,x}\hat{r}_{ij,z} \\ \gamma_{ij}\hat{r}_{ij,y}\hat{r}_{ij,x} & \beta_{ij}+\gamma_{ij}\hat{r}_{ij,y}^2 & \gamma_{ij}\hat{r}_{ij,y}\hat{r}_{ij,z} \\ \gamma_{ij}\hat{r}_{ij,z}\hat{r}_{ij,x} & \gamma_{ij}\hat{r}_{ij,z}\hat{r}_{ij,y} & \beta_{ij}+\gamma_{ij}\hat{r}_{ij,z}^2 \end{bmatrix} \tag{12}$$

where $\hat{r}_{ij,\alpha} = \dfrac{r_{ij,\alpha}}{r_{ij}}$ $(\alpha = x, y, z)$ \hfill (13)

$$C_{ij} = -\frac{k_0^2}{4\pi\varepsilon_0}\frac{e^{ik_0 r_{ij}}}{r_{ij}} \tag{14}$$

$$\beta_{ij} = \left[1 - \frac{1}{(k_0 r_{ij})^2} + \frac{i}{k_0 r_{ij}}\right] \tag{15}$$

$$\gamma_{ij} = -\left[1 - \frac{3}{(k_0 r_{ij})^2} + \frac{3i}{k_0 r_{ij}}\right] \tag{16}$$

In Eqs. (13) to (16), $r_{ij}$ is the magnitude of the distance vector between subvolumes $i$ and $j$. When $i = j$, the submatrix $\overline{\overline{\mathbf{A}}}_{ii}$ represents the self-interaction of subvolume $i$ in the absence of surface and its expression is given by $(1/\alpha_i)\overline{\overline{\mathbf{I}}}$.

The term $\overline{\overline{\mathbf{R}}}$ in Eq. (11) is the $3N$ by $3N$ deterministic reflection-interaction matrix that contains submatrices $\overline{\overline{\mathbf{R}}}_{ij}$ representing the interaction between subvolumes $i$ and $j$ due to reflection by the surface. A submatrix $\overline{\overline{\mathbf{R}}}_{ij}$ is obtained from the reflection DGF [35,46]:

$$\overline{\overline{\mathbf{R}}}_{ij} = C_{I,ij} \frac{\varepsilon_1 - 1}{\varepsilon_1 + 1} \begin{bmatrix} -(\beta_{I,ij} + \gamma_{I,ij}\hat{r}_{I,ij,x}^2) & -\gamma_{I,ij}\hat{r}_{I,ij,x}\hat{r}_{I,ij,y} & \gamma_{I,ij}\hat{r}_{I,ij,x}\hat{r}_{I,ij,z} \\ -\gamma_{I,ij}\hat{r}_{I,ij,x}\hat{r}_{I,ij,y} & -(\beta_{I,ij} + \gamma_{I,ij}\hat{r}_{I,ij,y}^2) & \gamma_{I,ij}\hat{r}_{I,ij,y}\hat{r}_{I,ij,z} \\ -\gamma_{I,ij}\hat{r}_{I,ij,x}\hat{r}_{I,ij,z} & -\gamma_{I,ij}\hat{r}_{I,ij,y}\hat{r}_{I,ij,z} & (\beta_{I,ij} + \gamma_{I,ij}\hat{r}_{I,ij,z}^2) \end{bmatrix}$$
$$-\frac{1}{4\pi\varepsilon_0} \begin{bmatrix} \hat{\rho}_{ij,x}^2 I_\rho^H - \hat{\rho}_{ij,y}^2 I_\varphi^H & \hat{\rho}_{ij,x}\hat{\rho}_{ij,y}(I_\rho^H + I_\varphi^H) & \hat{\rho}_{ij,x} I_\rho^V \\ \hat{\rho}_{ij,x}\hat{\rho}_{ij,y}(I_\rho^H + I_\varphi^H) & \hat{\rho}_{ij,y}^2 I_\rho^H - \hat{\rho}_{ij,x}^2 I_\varphi^H & \hat{\rho}_{ij,y} I_\rho^V \\ -\hat{\rho}_{ij,x} I_\rho^V & -\hat{\rho}_{ij,y} I_\rho^V & I_z^V \end{bmatrix} \tag{17}$$

where the subscripts $x$, $y$, and $z$ indicate vector components in Cartesian coordinates, while the subscripts $\rho$, $\varphi$, and $z$ refer to vector components in cylindrical coordinates. The parameters $\hat{r}_{I,ij,\alpha}$, $C_{I,ij}$, $\beta_{I,ij}$, and $\gamma_{I,ij}$ are defined in the same manner as in Eqs. (13) to (16), except that $r_{ij}$ is replaced by $r_{I,ij}$, which corresponds to the distance between subvolume $i$ and the image of subvolume $j$ within the surface. In Eq. (17), $\hat{\rho}_{ij,\alpha} = \frac{\rho_{ij,\alpha}}{|\boldsymbol{\rho}_{ij}|}$ ($\alpha = x, y$), where $\boldsymbol{\rho}_{ij}$ is the distance vector between

subvolumes $i$ and $j$ along the surface ($\boldsymbol{\rho}_{ij} = (x_i - x_j)\hat{\mathbf{x}} + (y_i - y_j)\hat{\mathbf{y}}$). The terms $I_\rho^V$, $I_z^V$, $I_\rho^H$, and $I_\varphi^H$ are defined as:

$$I_\rho^V = \frac{\partial^2}{\partial\rho\partial z} k_1^2 V'_{00} \tag{18}$$

$$I_z^V = \left(\frac{\partial^2}{\partial z^2} + k_0^2\right) k_1^2 V'_{00} \tag{19}$$

$$I_\rho^H = \frac{\partial^2}{\partial\rho^2} k_0^2 V'_{00} + k_0^2 U'_{00} \tag{20}$$

$$I_\varphi^H = -\frac{1}{\rho_{ij}} \frac{\partial}{\partial\rho} k_0^2 V'_{00} - k_0^2 U'_{00} \tag{21}$$

where $k_1$ is the magnitude of the wavevector in $V_1$, while $V'_{00}$ and $U'_{00}$ are the Sommerfeld integrals given by:

$$V'_{00} = 2i\int_0^\infty \left[\frac{1}{k_1^2 k_{z0} + k_0^2 k_{z1}} - \frac{1}{k_{z0}(k_1^2 + k_0^2)}\right] e^{ik_{z0}(z_i+z_j)} J_0(k_\rho \rho_{ij}) k_\rho dk_\rho \tag{22}$$

$$U'_{00} = 2i\int_0^\infty \left[\frac{1}{k_{z0} + k_{z1}} - \frac{k_0^2}{k_{z0}(k_1^2 + k_0^2)}\right] e^{ik_{z0}(z_i+z_j)} J_0(k_\rho \rho_{ij}) k_\rho dk_\rho \tag{23}$$

In Eqs. (22) and (23), $J_0$ is the zeroth-order Bessel function of the first kind, while $k_\rho$ and $k_{zj}$ are the wavevector components parallel and perpendicular to the surface, respectively. Note that the parallel component of the wavevector is a complex number, such that the Sommerfeld integrals are evaluated in the complex plane [46]. The $z$-components of the position vectors (i.e., $z_i$ and $z_j$) are calculated relative to the surface. Evaluation of the Sommerfeld integrals involves complex integration of multivalued functions $k_{zj}$ (= $(k_j^2 - k_\rho^2)^{1/2}$). Multivalued functions are represented in the complex plane via Riemann surfaces [47]. In order to perform the integration of a multivalued function, it is necessary to perform branch cuts such that the problem reduces to the integration of a single valued function on a single branch of the Riemann surface. Once branch cuts are performed, it is necessary to define the path of the integration. The integration path should avoid the poles of the function to be integrated and should not cross branch cuts in order to stay on a single branch of the Riemann surface. The integration path is not unique and is selected to ensure a fast convergence of the function to be integrated. In this work, the techniques and FORTRAN subroutines developed by Lager and Lytle [48,49] are used for calculating the Sommerfeld integrals.

The submatrix $\bar{\bar{\mathbf{R}}}_{ij}$ given by Eq. (17) represents the electric field intercepted by subvolume $i$ due to emission by subvolume $j$ after reflection at the surface. Mathematically, spherical waves emitted by subvolume $j$ are expressed using Eqs. (22) and (23) as the product of cylindrical waves propagating parallel to the surface (Bessel function) and plane waves propagating along the $z$-direction (exponential term). Only the plane wave component interacts with the surface. In the static limit or when the surface is a perfect electric conductor, $\bar{\bar{\mathbf{R}}}_{ij}$ can be obtained from the image theory using the direct interaction between subvolume $i$ and the image of subvolume $j$, with dipole moment $\frac{\varepsilon_1 - 1}{\varepsilon_1 + 1}\left(-p_{jx}\hat{\mathbf{x}} - p_{jy}\hat{\mathbf{y}} + p_{jz}\hat{\mathbf{z}}\right)$, within the surface [50]. The image contribution corresponds to the first term on the right-hand side of Eq. (17). In this paper, the general case where $\bar{\bar{\mathbf{R}}}_{ij}$ is calculated from Eq. (17) is considered.

The net spectral heat rate between the surface and the objects is defined as $\langle Q_{net,\omega} \rangle = \sum_i \langle Q_{abs,\omega,1i} \rangle - \sum_i \langle Q_{abs,\omega,i1} \rangle$, where $\langle Q_{abs,\omega,1i} \rangle$ is the spectral power absorbed by subvolume $i$ due to thermal emission by the surface and vice-versa for $\langle Q_{abs,\omega,i1} \rangle$, while $\langle\,\rangle$ denotes a time average. Using reciprocity, the net spectral heat rate can be expressed solely in terms of $\langle Q_{abs,\omega,1i} \rangle$:

$$\langle Q_{net,\omega} \rangle = \sum_i \langle Q_{abs,\omega,1i} \rangle \left[ \frac{\Theta(\omega,T_{2i})}{\Theta(\omega,T_1)} - 1 \right] \tag{24}$$

where $T_{2i}$ is the temperature of subvolume $i$. The power absorbed $\langle Q_{abs,\omega,1i} \rangle$ is calculated from the induced dipole moments as follows:

$$\langle Q_{abs,\omega,1i} \rangle = \frac{\omega}{2}\left( \mathrm{Im}[(\alpha_i^{-1})^*] - \frac{2}{3}k_0^3 \right) \mathrm{tr}\langle \mathbf{p}_i^{ind} \otimes \mathbf{p}_i^{ind} \rangle \tag{25}$$

where ergodicity is assumed [51]. Note that when calculating the power absorbed, it is assumed that the objects described by $V_2$ are non-emitting and purely absorbing ($T_{2i} = 0$ K). Yet, thermal emission by $V_2$ is accounted for by capitalizing on reciprocity, as shown by Eq. (24). Therefore, $\mathbf{p}_i^{fl} = \mathbf{0}$ and $\mathbf{p}_i = \mathbf{p}_i^{ind}$ for all subvolumes contained in $V_2$. The trace of the autocorrelation function of the induced dipole moments in Eq. (25) is obtained directly from the system of equations (11):

$$\langle \bar{\mathbf{P}} \otimes \bar{\mathbf{P}} \rangle = (\bar{\bar{\mathbf{A}}} + \bar{\bar{\mathbf{R}}})^{-1} \cdot \langle (\bar{\mathbf{E}}^{inc} \otimes \bar{\mathbf{E}}^{inc}) + (\bar{\mathbf{E}}^{sur} \otimes \bar{\mathbf{E}}^{sur}) \rangle \cdot \left( (\bar{\bar{\mathbf{A}}} + \bar{\bar{\mathbf{R}}})^{-1} \right)^\dagger \tag{26}$$

where the superscript † indicates the Hermitian operator defined as the conjugate transpose. The ensemble average of the spatial correlation function of the surface fields in subvolumes $i$ and $j$ is calculated as [52]:

$$\langle \mathbf{E}_i^{sur} \otimes \mathbf{E}_j^{sur} \rangle = \frac{\varepsilon_0 \mu_0^2 \omega^3 \varepsilon_1''}{8\pi^3} \Theta(\omega, T_1)$$
$$\times \int_0^\infty \int_0^{2\pi} \frac{k_\rho}{|k_{z1}|^2 k_{z1}''} e^{i(k_{z0}z_i - k_{z0}^* z_j)} e^{i\mathbf{k}_\rho \cdot \mathbf{\rho}_{ij}} \left[ |t_{10}^s|^2 (\hat{\mathbf{s}} \cdot \hat{\mathbf{s}})(\hat{\mathbf{s}} \otimes \hat{\mathbf{s}}) + |t_{10}^p|^2 (\hat{\mathbf{p}}_1 \cdot \hat{\mathbf{p}}_1^*)(\hat{\mathbf{p}}_0 \otimes \hat{\mathbf{p}}_0^*) \right] d\varphi dk_\rho \quad (27)$$

where $k_{z1}''$ is the imaginary part of $k_{z1}$, $t_{10}^s$ and $t_{10}^p$ are the Fresnel transmission coefficients for transverse electric (TE) and transverse magnetic (TM) polarizations, and the azimuthal angle $\varphi$ is measured between the parallel component of the wavevector and the $x$-axis. The terms $\hat{\mathbf{s}}$ and $\hat{\mathbf{p}}$ are unit vectors oriented along the TE and TM polarizations, respectively [44]. In Eq. (27), the parallel component of the wavevector $k_\rho$ is a real number since the electric field does not vary along the $x$- and $y$-directions for an emitting infinite surface.

Once the dipole moment correlation matrix is computed with Eq. (26), the power absorbed in subvolume $i$, $\langle Q_{abs,\omega,1i} \rangle$, and the net heat rate between the objects and the surface, $\langle Q_{net,\omega} \rangle$, are respectively calculated with Eqs. (25) and (24).

As described in this section, the T-DDA formalism only involves numerical approximations. Therefore, the T-DDA can be considered as numerically exact since the results obtained from this method converge to the exact solution in the limit that $N \to \infty$. In the next section, the framework is verified against the exact solution of the heat rate between a sphere and a surface.

### III. VERIFICATION OF THE T-DDA WITH SURFACE INTERACTION

The T-DDA is verified by comparison against the exact solution of the heat rate between a 1.6-µm-diameter sphere and a surface [10]. The surface and the sphere are both made of silica and are maintained at temperatures $T_1 = 300$ K and $T_2 = 400$ K, respectively. The dielectric function of silica has been taken from Ref. [53] and is shown in Fig. 3. It is assumed that there is no incident electric field. Figure 4 shows the net spectral heat rate obtained from the exact solution and the T-DDA for separation gaps $d$ of 100 nm and 100 µm.

The convergence of the T-DDA depends strongly on the dielectric function of the discretized object [18]. As the dielectric function increases, the wavelength and the decay length of the electric field (skin depth) inside the object shrinks. As such, the subvolume size resulting in a converged solution decreases as the dielectric function increases. Additionally, a large dielectric function negatively affects the T-DDA convergence by amplifying the shape error [18]. Since the dielectric function of silica varies significantly with the frequency in the infrared band (see Fig. 3), a frequency-dependent nonuniform discretization was used for calculating the net spectral heat rate. The number of subvolumes employed for discretizing the sphere varied between 11536 and 33552 depending on the frequency. The computational time required for performing the simulations depends on the number of subvolumes, the separation gap and the discretization scheme (uniform or nonuniform discretization). For the case shown in Fig. 4, the computational time varied between 10 and 250 service units (core-hours) per frequency using Intel Xeon E5-2670 processors with a processing speed of 2.60 GHz.

It can be seen in Fig. 4 that the T-DDA and exact results are in excellent agreement. The locations of the resonances and their magnitudes are predicted accurately via the T-DDA. The small discrepancy observed for frequencies ranging from 0.1300 eV to 0.1375 eV is due to the fact that the dielectric function of silica is large within that spectral band. A better accuracy could be obtained by employing a larger number of subvolumes, since the accuracy of the T-DDA increases as the subvolume size decreases [18]. Yet, increasing the number of subvolumes within the 0.1300-0.1375 eV spectral band is not necessary as its contribution to the net total heat rate is negligible. The satisfactory results obtained for the case of a sphere, which is one of the most difficult shapes to model with a cubical lattice, demonstrates that the T-DDA can accurately be used for modeling arbitrarily-shaped objects. The T-DDA with surface interaction is applied next to near-field radiative heat transfer between a complex-shaped probe and a surface.

## IV. NEAR-FIELD RADIATIVE HEAT TRANSFER BETWEEN A PROBE AND A SURFACE

The framework described in Section II is used hereafter to analyze radiative heat transfer between a probe and a surface. The probe geometry consists of an assembly of a rectangular cuboid, a conical frustum and a cylinder (see Fig. 5). The base and the height of the cuboid have dimensions of 57.8 nm and 288.9 nm, respectively. The conical frustum has a height of 3.872 μm, and the diameters of its lower and upper bases are 115.6 nm and 1.16 μm, respectively. The cylinder has a height of 809.0 nm and a diameter of 1.16 μm. The probe and the surface are both made of silica. In all cases, the surface is at a temperature $T_1$ = 300 K while the probe is at $T_2$ = 400 K. For simplicity, it is assumed that there is no incident electric field. For far-field simulations ($d$ = 100 μm), 11113 uniform subvolumes were used to discretize the probe while 13111 nonuniform subvolumes were employed for near-field simulations ($d$ = 10 nm). Increasing the number of subvolumes beyond these values did not affect the results. The computational time for the probe-surface configuration varied between 10 and 70 service units per frequency.

### A. Spectral distribution of heat rate and near-field regimes

Figure 6(a) shows the net spectral heat rate for separation gaps $d$ between the probe and the surface of 10 nm, 100 nm and 100 μm. For purpose of comparison, the net spectral heat rate for a sphere of same material, volume (diameter of 1.6 μm) and temperature as the probe is reported in Fig. 6(b). The heat rate profiles exhibit low-frequency (~ 0.06 eV) and high-frequency (~ 0.14 eV) resonances due to surface phonon-polaritons (SPhPs) and localized surface phonons (LSPhs). At separation gaps of 100 nm and 100 µm, both the low- and high-frequency resonances of the probe-surface heat rate are split into two modes, while this splitting is not observed for a 10-nm-thick gap as well as in the sphere-surface configuration. The origin of these resonances can be explained by first considering the near-field thermal spectrum of the surface in the absence of object, characterized by the energy density, as shown in Fig. 7 for distances of 10 nm, 100 nm and 100 µm. In the near field (10 nm and 100 nm), low- and high-frequency resonances are observed at 0.0613 eV and 0.1435 eV due to thermal excitation of SPhPs. When losses are small ($\varepsilon_1'' \to 0$), SPhPs are resonantly excited at a flat material-vacuum interface when the real part of the dielectric function $\varepsilon_1'$ equals -1 [45]. Here, the high-frequency resonance occurs when $\varepsilon_1' = -1$, while the low-frequency SPhP mode arises when $\varepsilon_1'$ is equal to -0.83 due to non-negligible losses. In the far field (100 µm), SPhPs lead to low thermal emission resulting in local minima in the energy density profile. Yet, when an object is located at a distance of 100 µm above the surface, low- and high-frequency resonances arise due to LSPhs supported by the sphere and the probe. The electric dipole approximation can be used for estimating these LSPh modes. The power absorbed by an electric dipole is proportional to Im($\alpha_j$), where $\alpha_j$ ($j = x, y, z$) is the dipole polarizability tensor given by [54,55]:

$$\alpha_j = \frac{4\pi}{3} \varepsilon_0 a_x a_y a_z \frac{\varepsilon_2 - 1}{1 + L_j (\varepsilon_2 - 1)} \tag{28}$$

where $a_x$, $a_y$ and $a_z$ are the dimensions of the dipole along the $x$-, $y$- and $z$-directions. The geometrical factors $L_j$, determined solely from the dipole geometry, satisfy $\sum_{j=x,y,z} L_j = 1$ and $L_j \geq 0$.

For a spherical dipole, $L_x = L_y = L_z = 1/3$ such that Eq. (28) reduces to the Clausius-Mossotti polarizability with resonant enhancement when $|\varepsilon_2 + 2|$ is minimum. In this limit, LSPh resonances are predicted at frequencies of 0.0605 eV and 0.1410 eV, which is in good agreement with the resonances of the heat rate for the case of a sphere located 100 μm above the surface (0.0605 eV and 0.1400 eV). For the probe, LSPh resonances can be estimated by considering a prolate spheroidal dipole having a major axis $a_z$ equal to the probe length of 4.97 μm. The minor axes $a_x$ and $a_y$ are the same and are equal to 321 nm such that the spheroid and the probe have the same volume. For these dimensions, the geometrical factors needed to calculate the polarizability tensor are $L_x = L_y = 0.495$ and $L_z = 0.010$. Thus, resonant enhancement due to LSPhs for the case of a prolate spheroidal electric dipole occurs along the major axis and minor axes when $|\varepsilon_2 + 99|$ and $|\varepsilon_2 + 1.02|$ are minimum, respectively. In this limit, four resonant modes along the minor axes are predicted at frequencies of 0.0575 eV, 0.0615 eV, 0.1325 eV and 0.1450 eV. These predictions are in good agreement with the resonant modes of the heat rate profile for a probe located 100 μm above the surface.

As the gap decreases to 100 nm and 10 nm, the heat rate for both the probe and sphere cases increases due to the additional contribution of evanescent modes, and particularly due to SPhPs supported by the surface. In the near field, the sphere is optically thick and its diameter is much larger than the gap distance, such that heat transfer can be approximated as a summation of local heat rates between two parallel surfaces with varying gap thicknesses (proximity approximation) [15,17]. Consequently, the resonant frequencies of the near-field heat rate profiles are essentially the same as the SPhP resonant frequencies of the surface. In the proximity approximation limit, the total near-field conductance, which is proportional to the net total heat rate, is calculated as:

$$G(T) = \lim_{\delta T \to 0} \frac{\langle Q_{net} \rangle}{\delta T} = \int_A h(\tilde{d},T) dA \qquad (29)$$

where $\tilde{d}$ is the local distance between two parallel surfaces and $h$ is the heat transfer coefficient in the two-surface configuration that follows a $\tilde{d}^{-2}$ power law [45]. For a sphere and a surface, Eq. (29) results in a $d^{-1}$ power law for small gaps up to approximately 100 nm [15]. This is shown in Fig. 8, where the sphere-surface total heat rate is plotted as a function of the separation gap (10 nm to 100 nm).

The probe-surface spectral heat rate for a 100-nm-thick gap is similar to the far-field profile, where both low- and high-frequency resonance splitting mediated by LSPhs along the minor axis of the probe is observed. However, when the gap reduces to 10 nm, which is smaller than the probe tip size of 57.8 nm, resonance splitting does not occur. Instead, the resonances are aligned with those of the sphere-surface configuration in the near field, and thus essentially correspond to SPhP modes of the surface. Here, heat transfer is dominated by SPhPs with penetration depth approximately equal to the gap size and thus smaller than the probe tip size [56,57]. Consequently, the heat rate between the probe and the surface can also be estimated using the proximity approximation. For a probe with a flat tip, the total heat rate is expected to follow a $d^{-2}$ power law in the limit that $d \to 0$ since $\tilde{d}$ in Eq. (29) is independent of the surface area $A$. Near a 10-nm-thick gap, Fig. 8 shows that the heat rate varies as $d^{-1.5}$ while a $d^{-0.3}$ power law is observed around a gap size of 100 nm. It can be seen in the inset of Fig. 8 that the heat rate decays as $d^{-1.77}$ near a gap size of 6 nm, such that the $d^{-2}$ regime is expected to arise at an extremely small gap where the validity of fluctuational electrodynamics is questionable. From these results, it is concluded that the $d^{-2}$ regime is reached when the probe tip size is larger than the gap by more than one order of magnitude. Note that for a spheroid and a cone above a surface, the integration of Eq. (29) leads to $d^{-1}$ and $\log(d^{-1})$ power laws, such that these geometries do not represent well the heat rate between the probe and the surface in the framework of the proximity approximation.

To summarize, the decay rate of the probe-surface near-field heat transfer decreases as the gap thickness increases. When the gap size is much smaller than the probe tip size, SPhP mediated heat transfer between the probe and the surface occurs such that the heat rate follows a power law approaching $d^{-2}$. For gap sizes of the same order of magnitude as or larger than the probe tip size, the near-field heat rate is mediated by coupled SPhP and LSPh (along the minor axis of the probe) modes, resulting in a decay rate of $d^{-0.3}$. In the far field, heat transfer between the probe and the surface is dominated by LSPhs. The spatial distribution of volumetric heat rate, normalized by its maximum value, at the low-frequency resonance is plotted for SPhP (10 nm), coupled SPhP-LSPh (100 nm) and LSPh (100 μm) mediated heat transfer in Fig. 9. For 100 nm and 100 μm gaps, the first mode of the low-frequency resonance is considered. Note that the spatial distribution of volumetric heat rate is shown for a cross-section parallel to the *y-z* plane passing through the central axis of the probe. It is clear from Fig. 9 that as the contribution of SPhPs increases, the heat absorbed by the probe is essentially concentrated at its tip. As a final remark, note that the peak observed at 0.10 eV in Fig. 6 is due to a local maximum in the imaginary part of the dielectric function of silica leading to increased radiation absorption by the object and enhanced contribution of frustrated modes in the near field. As the distance between the object and the surface decreases, the contribution from this peak decreases and becomes essentially negligible at a gap distance of 10 nm where the heat rate is dominated by SPhPs.

**B. Validity of the spheroidal electric dipole approximation for modeling near-field radiative heat transfer between a probe and a surface**

The validity of the spheroidal electric dipole model for approximating near-field radiative heat transfer between a probe and a surface is analyzed. Note that the spherical electric dipole approximation is not considered, as it cannot predicts the splitting of the low- and high-frequency resonances of the spectral heat rate profile (see Fig. 6(a)). The net spectral heat rate between an electric dipole and a surface is calculated as [19,21]:

$$\langle Q_{net,\omega} \rangle = \frac{2\omega^4 \mu_0^2 \varepsilon_0 \varepsilon_1''}{\pi} [\Theta(\omega,T_2) - \Theta(\omega,T_1)] \sum_{j=x,y,z} \left( \text{Im}(\alpha_j) \sum_{k=x,y,z} \int_{V_1} |G_{jk}^T(\mathbf{r}_2,\mathbf{r}',\omega)|^2 d^3\mathbf{r}' \right) \quad (30)$$

where $\alpha_j$ is the polarizability tensor given by Eq. (28) and $\mathbf{r}_2$ is the distance between the centroid of the dipole and the surface. Figure 10 shows the net spectral heat rate for the prolate spheroidal dipole and probe discussed in Section IV.A at separation gaps $d$ of 10 nm and 100 nm, while the net total heat rate is provided in Fig. 11 for gaps ranging from 10 nm to 500 nm. The spheroidal electric dipole model predicts low- and high-frequency resonance splitting regardless of the gap size. As discussed previously, these four resonances are due to LSPhs associated with the minor axes of the prolate spheroidal dipole and are thus independent of the gap size. It can be seen in Fig. 11 that the total near-field heat rate in the spheroidal dipole approximation is a weak function of the gap size. This is due to the fact that the dipole centroid is located at a distance $d + a_z/2$ ($a_z$ = 4.97 µm) above the surface, such that variations of $d$ by a few tens to a few hundreds of nanometers do not significantly affect heat transfer.

According to Ref. [21], the spheroidal dipole model is expected to provide reliable results when the wavelength $\lambda$ and the gap $d$ is larger than $a_{max} = \max\{a_x,a_y,a_z\}$. Figure 11 however suggests that starting at a gap size of approximately 70 nm, where the probe tip size is smaller than the gap size, the spheroidal dipole model approximates reasonably well the heat rate between a probe and a surface. In addition, as discussed in Section IV.A, the resonant modes of the probe-surface heat rate are well predicted by the spheroidal dipole approximation when heat transfer is mediated by coupled SPhPs-LSPhs. For a spheroidal dipole, a more appropriate criterion should require that the gap thickness $d$ be much larger than the radius of curvature $R$ of the spheroidal dipole tip facing the surface ($R = 10.4$ nm). When $d \gg R$, multiple reflections between the probe and the surface can be ignored. This new criterion assessing the applicability of the spheroidal dipole approximation to model the heat rate between a probe and a surface is in line with the results observed in Figs. 10 and 11. Yet, the total heat rates obtained for the probe and the spheroidal dipole are not in perfect agreement, and some discrepancies can be observed between gaps of 100 nm and 500 nm. As seen in Fig. 10 at a gap size of 100 nm, the spheroidal dipole model overestimates the heat rate associated with the first mode of the high-frequency resonance (0.1325 eV). This can be explained by the fact that the material wavelength corresponding to that frequency is 3.93 µm, which is of the same order of magnitude as the major axis of the spheroid. In the SPhP regime, where the probe tip size is larger than the gap size, the spheroidal dipole model cannot be used for approximating heat transfer between a probe and a surface, both in terms of resonance and total heat rate predictions, since the criterion $d \gg R$ is not respected. For this case, the proximity approximation can be employed to estimate the heat rate and resonant modes in the probe-surface configuration.

## V. CONCLUSIONS

A general formalism for modeling near-field radiative heat transfer between arbitrarily-shaped objects and an infinite surface was proposed. The thermal discrete dipole approximation (T-DDA) was used to discretize the volume integral equation for the electric field derived from fluctuational electrodynamics, while the surface interactions were treated analytically using Sommerfeld's theory of electric dipole radiation above an infinite plane. The framework was verified against the exact solution of heat rate between a sphere and a surface, and was then applied to near-field radiative heat transfer between a complex-shaped probe and a surface both made of silica. The study revealed that when the probe tip size is much larger than the separation gap $d$, surface phonon-polariton (SPhP) mediated heat transfer occurs such that the resonances of the heat rate correspond essentially to those of a single surface while the total heat rate approaches a $d^{-2}$ power law as $d \to 0$. It was also found that coupled localized surface phonon (LSPh)-SPhP mediated heat transfer arises when the probe tip size is approximately equal to or smaller than the separation gap. In that case, the spectral heat rate exhibits four resonant modes due to LSPhs along the minor axis of the probe while the total heat rate in the near field convergences to a $d^{-0.3}$ regime. Finally, it was demonstrated that a prolate spheroidal electric dipole can approximate reasonably well near-field radiative heat transfer between a probe and a surface when the thermal wavelength is larger than the major axis of the spheroidal dipole and when the separation gap is much larger than the radius of curvature of the dipole tip facing the surface. The framework presented in this paper is not restricted to the probe-surface configuration, and can be applied to cases involving an arbitrary number of objects with various sizes and shapes above an infinite plane.

## ACKNOWLEDGMENTS

This work was sponsored by the US Army Research Office under Grant No. W911NF-14-1-0210. The authors also acknowledge the Center for High Performance Computing at the University of Utah for providing the computational resources used in this study.

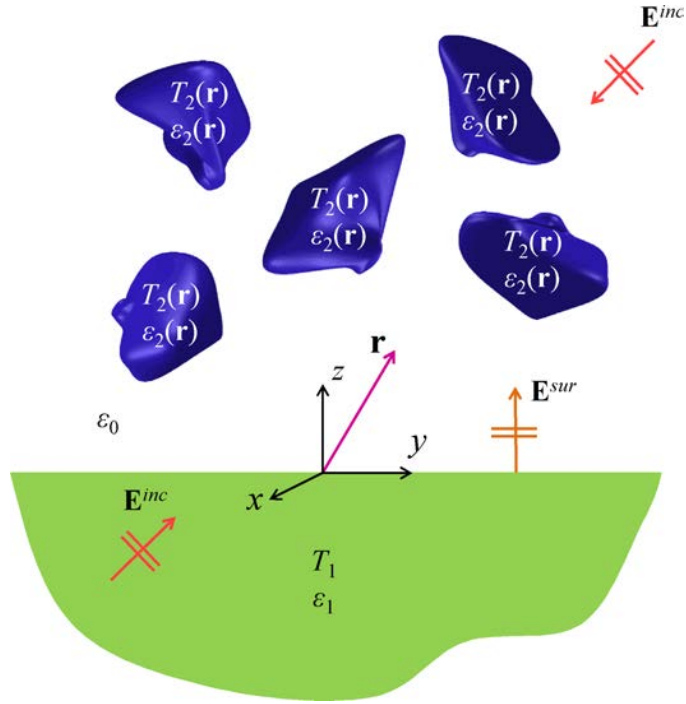

FIG. 1. (Color online) Schematic representation of the problem under consideration. Objects (medium 2) are submerged in vacuum (medium 0) above an infinite surface (medium 1). The incident electric field $\mathbf{E}^{inc}$ accounts for illumination by external sources, while the surface field $\mathbf{E}^{sur}$ is the electric field due to thermal emission by the surface.

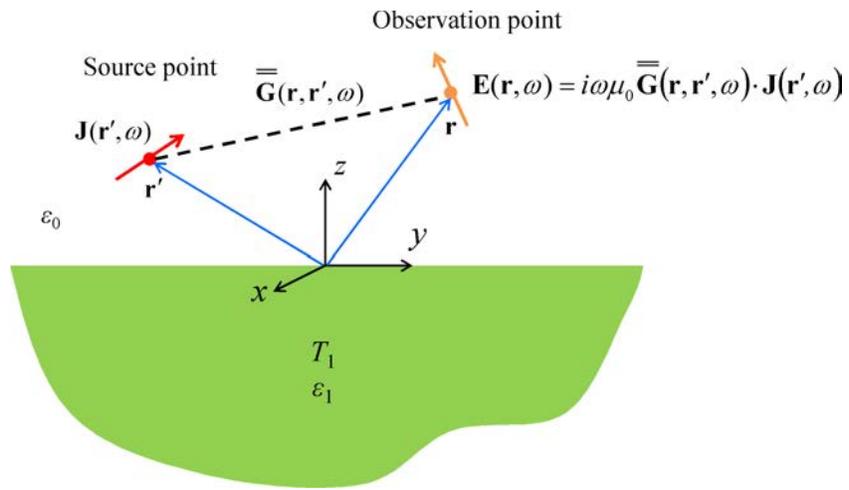

FIG. 2. (Color online) Dyadic Green's function (DGF) relating the electric field at point **r** to a source located at point **r′** in the presence of a surface.

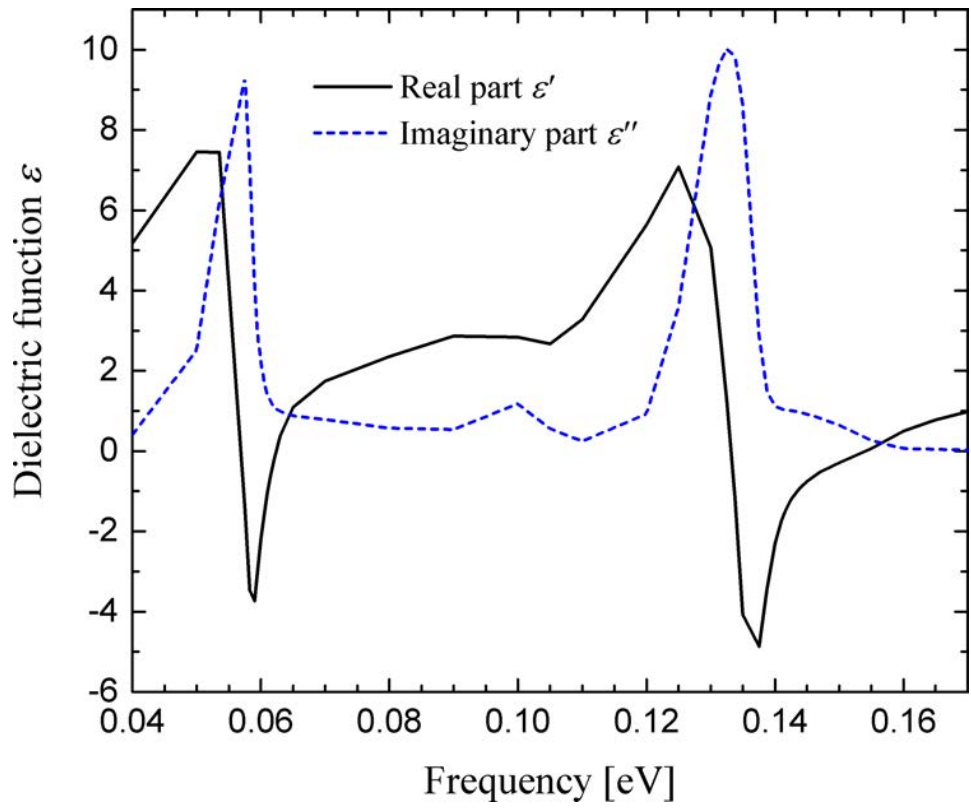

FIG. 3. (Color online) Dielectric function of silica obtained from Ref. [53].

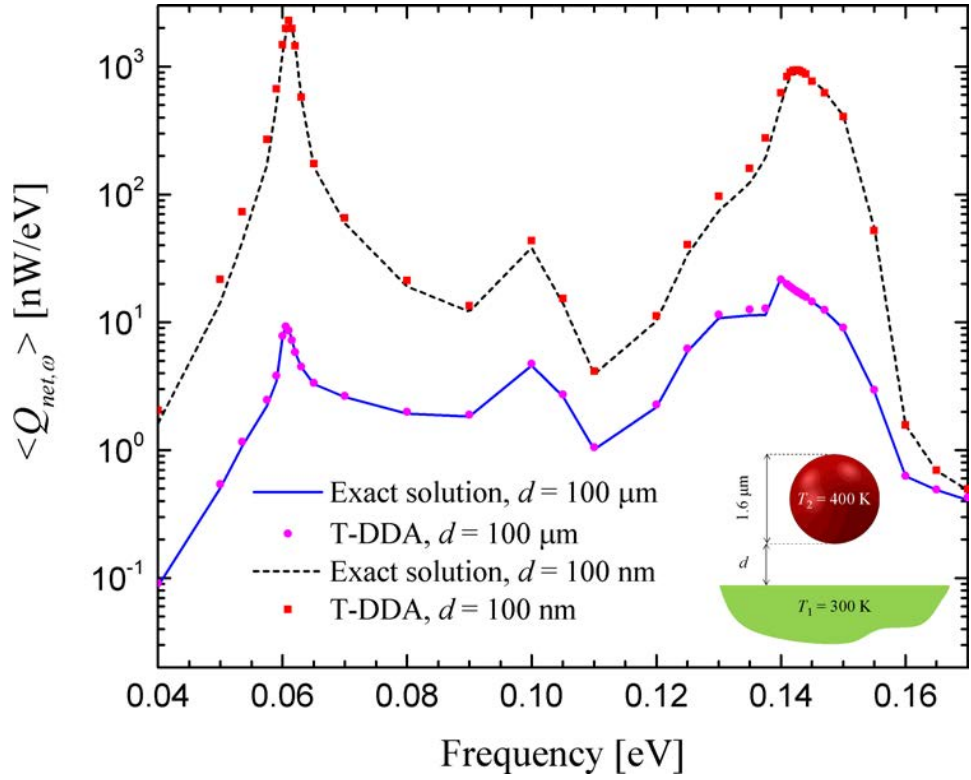

FIG. 4. (Color online) Net spectral heat rate between a sphere and a surface for separation gaps $d$ of 100 nm and 100 μm obtained with the T-DDA and the exact solution [10]. The sphere is at a temperature $T_2 = 400$ K, while the surface is at $T_1 = 300$ K. The sphere and the surface are made of silica, and the sphere diameter is 1.6 μm.

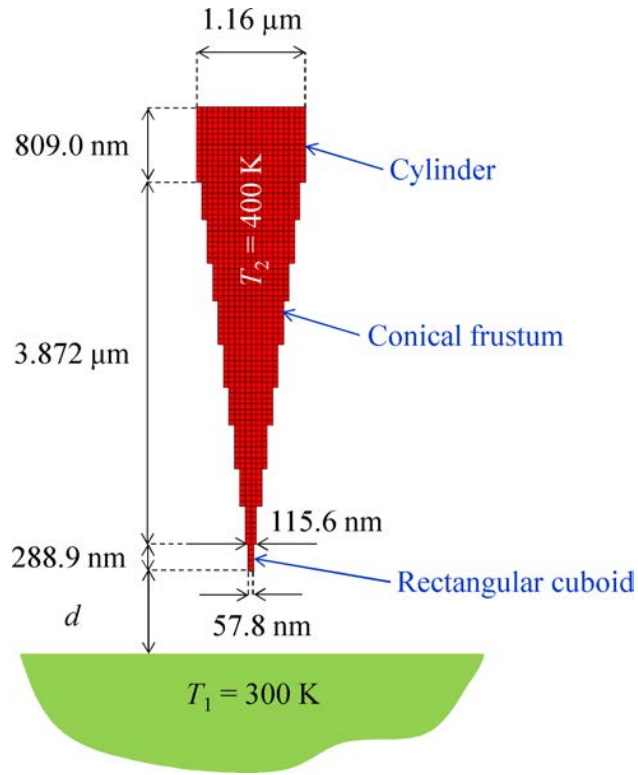

FIG. 5. (Color online) Radiative heat transfer between a probe and a surface separated by a gap of thickness *d*. The probe is modeled as an assembly of a rectangular cuboid, a conical frustum and a cylinder.

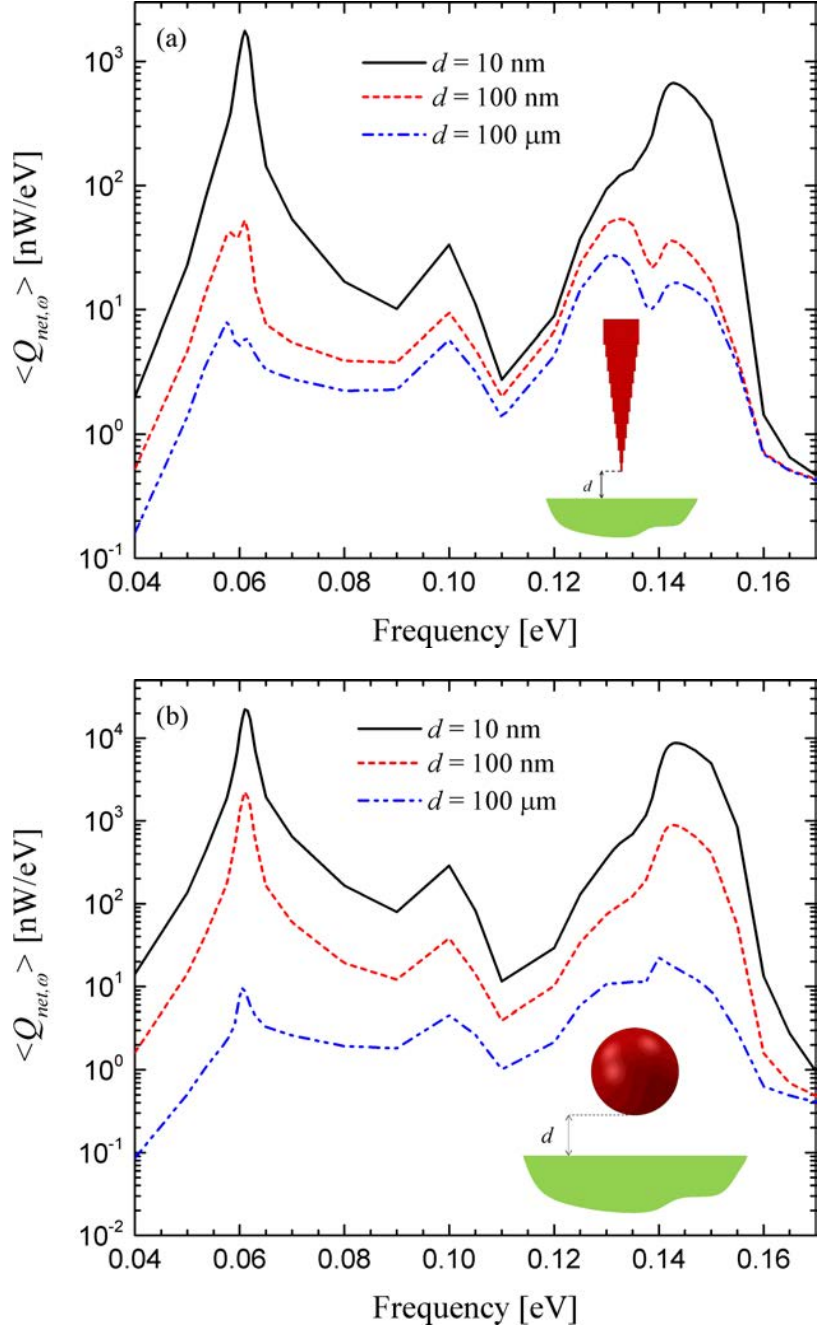

FIG. 6. (Color online) Net spectral heat rate between (a) a probe and a surface, and (b) a sphere and a surface for separation gaps $d$ of 10 nm, 100 nm and 100 μm. The sphere and the probe are at temperature $T_2 = 400$ K, while the surface is at temperature $T_1 = 300$ K. The sphere diameter is 1.6 μm, and the probe dimensions are shown in Fig. 5. The sphere, the probe and the surface are made of silica.

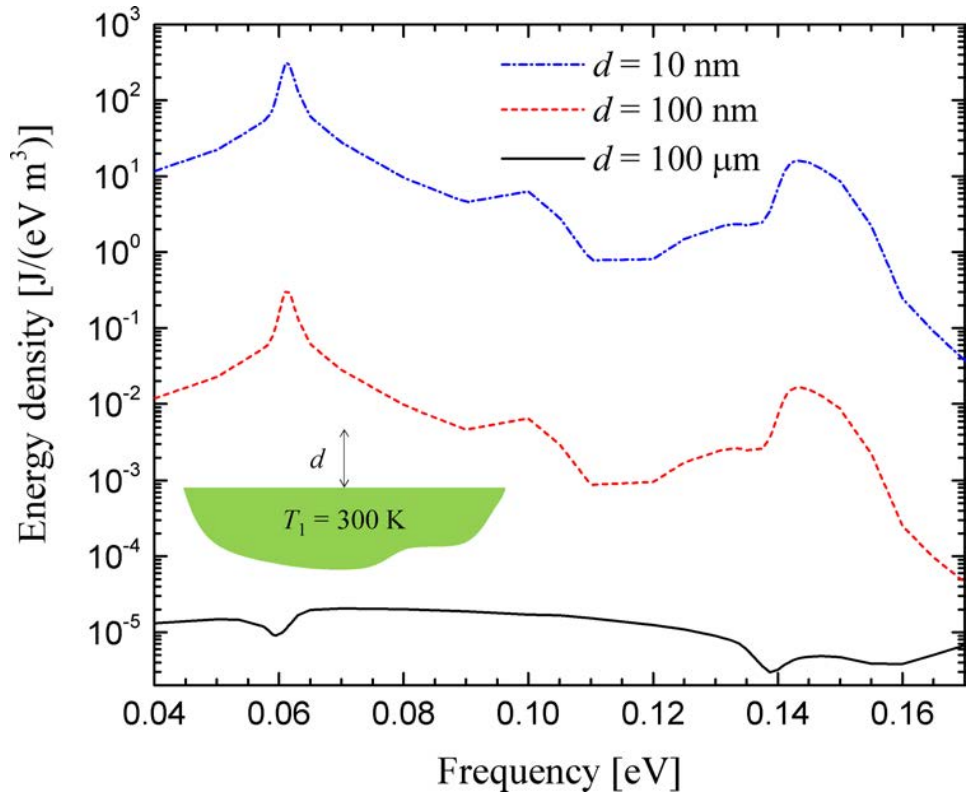

FIG. 7. (Color online) Spectral distribution of energy density at distances $d$ of 10 nm, 100 nm and 100 μm above a silica surface at a temperature $T_1 = 300$ K.

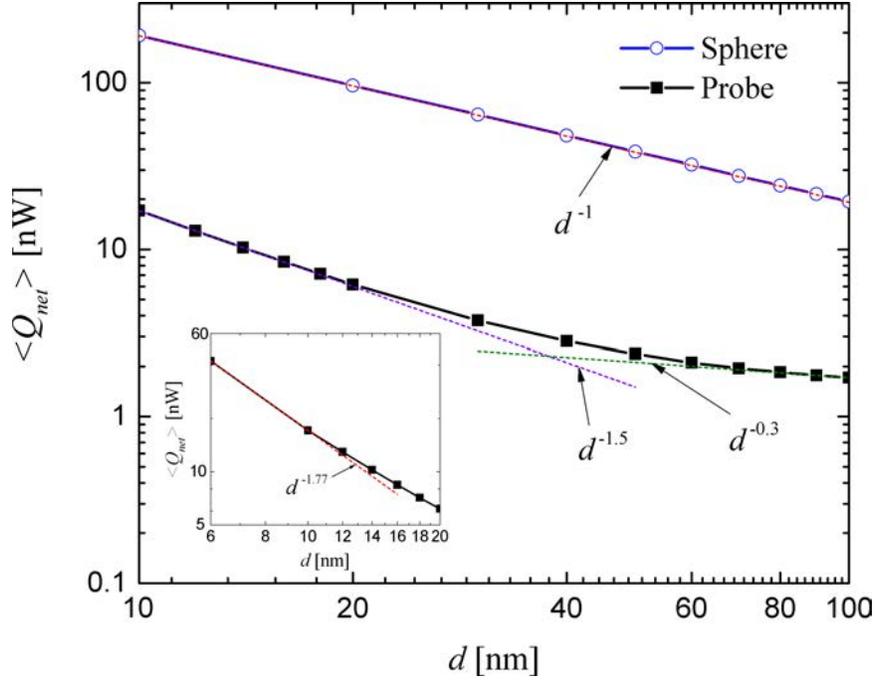

FIG. 8. (Color online) Net total heat rate in the near field as a function of the separation gap $d$ for the probe-surface and sphere-surface configurations. The sphere and the probe are at $T_2 = 400$ K, while the surface is at $T_1 = 300$ K. The sphere diameter is 1.6 μm, and the probe dimensions are shown in Fig. 5. The sphere, the probe and the surface are made of silica.

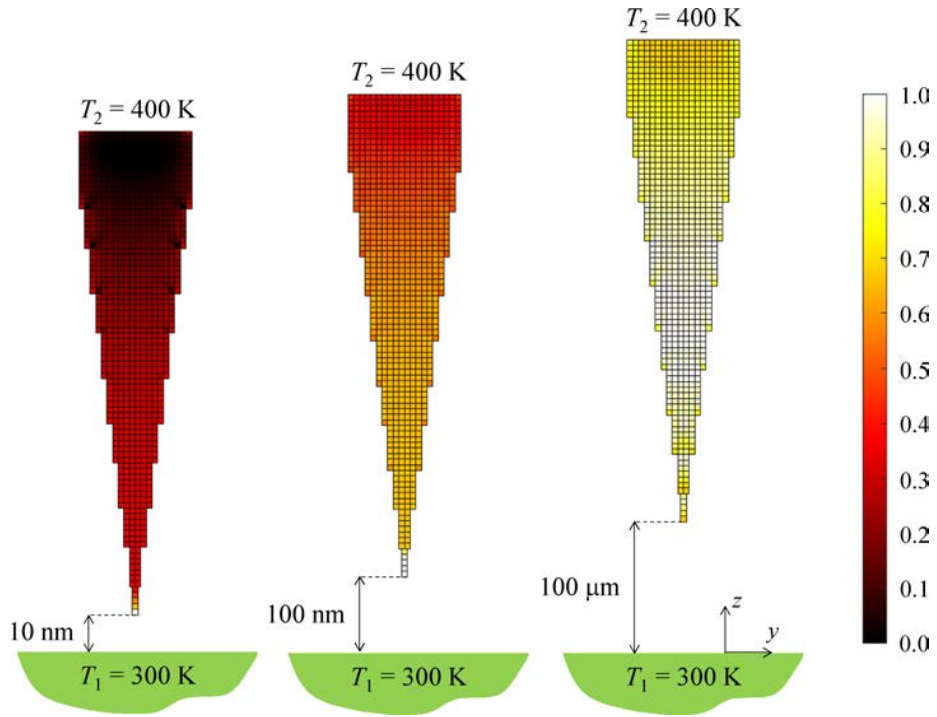

FIG. 9. (Color online) Spatial distribution of normalized volumetric heat rate within the probe at the low-frequency resonance for gap sizes $d$ of 10 nm, 100 nm and 100 μm. The probe and the surface are made of silica.

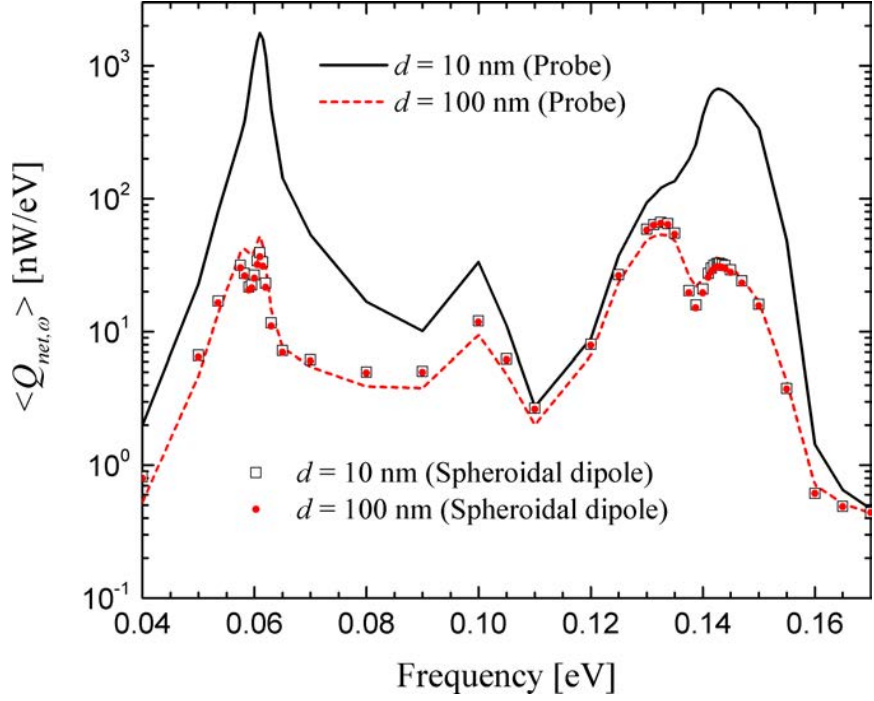

FIG. 10. (Color online) Net spectral heat rate between a probe and a surface for gap sizes $d$ of 10 nm and 100 nm. Results are compared against the spheroidal electric dipole model. The probe is at $T_2 = 400$ K, while the surface is at $T_1 = 300$ K. The probe and the surface are made of silica.

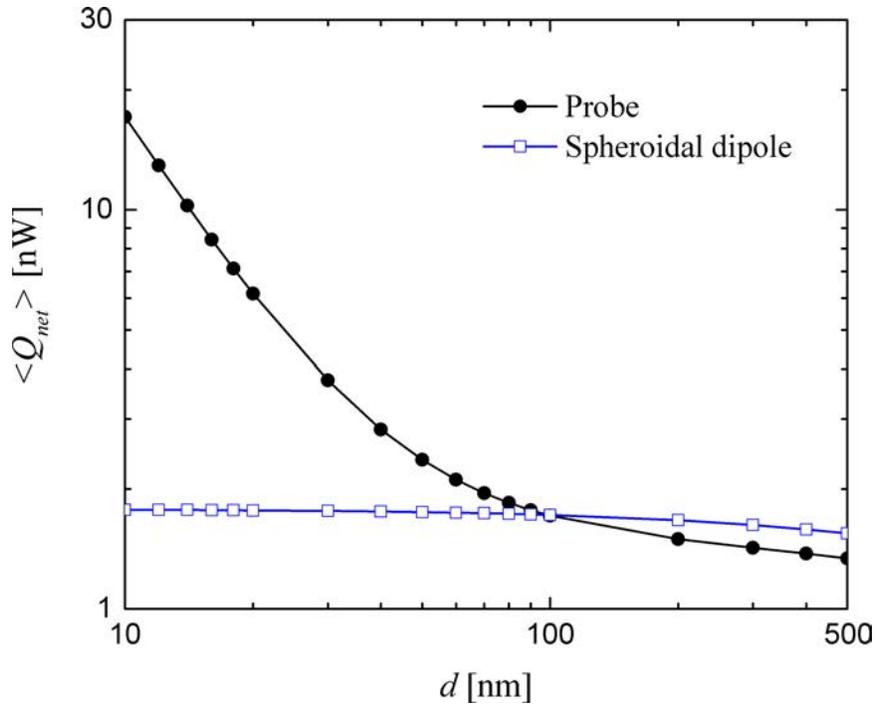

FIG. 11. (Color online) Net total heat rate between a probe and a surface as a function of the separation gap $d$. Results are compared against the spheroidal electric dipole model. The probe is at $T_2 = 400$ K, while the surface is at $T_1 = 300$ K. The probe and the surface are made of silica.